# Tunable Electronic, Optical and Transport Properties of Two Dimensional GaS Doped with Group II and Group IVa Elements

Munish Sharma

*Department of Physics, School of Basic and Applied Sciences, Maharaja Agrasen University, Atal Shiksha Kunj, Barotiwala, 174103, Himachal Pradesh (India)*

(May 16, 2019)

Corresponding Author

Email: munishsharmahpu@live.com

**Abstract**

In this present study, we systematically investigated the structural, electronic, optical and transport properties of pristine and group II, group IVa doped GaS monolayers using density functional theory (DFT). The strong formation energy suggests realization of group II & group IVa doped GaS. A semiconductor to metallic transition occurs in GaS monolayer with doping. Moreover, doped GaS monolayers have shown tunable optical properties. Doped GaS monolayers show optical activity in both ultraviolet (UV) as well as visible region. EEL spectra of GaS shift towards high energy region (red shift) with group II elements doping while no significant shift is observed for group IVa elements doping in both the polarizations. We found quantum conductance of $4G_0$ for doped GaS monolayer except for Be-doped GaS. The metallic character of doped GaS is clearly captured in tunneling current characteristics showing ohmic-like characteristics for doped GaS monolayer. This is due to finite density of states associated with S atoms of doped GaS. The doping extends the functionalities of pristine GaS with tunable properties suggesting doped GaS as a potential candidate for use in photo-diodes, photo-catalysts, photo-detectors and bio-sensing.

## 1. Introduction

The research in two dimensional materials has gained great attention in the last decade after discovery of graphene. A variety of 2D materials having unique physiochemical properties been discovered[1] and are being further explored for use in variety of fields including optoelectronics[2], molecular sensing[3], spintronics[4], catalysis[5], thermal energy management[6] etc. Gallium and Indium mono-chalcogenides are recent addition to family of 2D materials in hexagonal phase. The 2D Gallium and Indium mono-chalcogenides can be obtained by a similar mechanical exfoliation techniques as applied to graphene and transition metal di-chalcogenides (TMDs)[7]. This typical member is a quasi-two dimensional layered material have general formula MX, where M= Ga, In and X=S, Se. Unlike Transition metal chalcogenides, MX monolayers are formed by two covalently bonded double layers of Gallium atoms sandwiched between Sulfur atoms with $D_{3h}$ symmetry. The stacking arrangements of atoms results in different polytypes of macroscopic crystal[8].

Although, transition metal di-chalcogenides have already pushed many 2D materials like graphene and silicene on backbench due to their versatile properties ranging from insulator (e.g. $HfS_2$), semiconductors (e.g. $MoS_2$ and $WS_2$ etc.), semimetals (e.g. $WTe_2$, $TiSe_2$ etc.), to metals (e.g. $NbS_2$ and $VSe_2$)[9], but unusual 'Mexican Hat' like dispersion in valance band near Fermi level make MX monolayers unique for optoelectronic devices[10]. The ultra-thin MX layers are found to have large second harmonic generation and high Photoresponsibility indicating application in optoelectronics[11]. The fabrication of transistors based on GaS and GaSe ultrathin layers have already been reported with a good ON/OFF ratio and electron differential mobility[7, 11a].

In the recent years, variety of approaches has been adopted to tune the properties of novel MX layers. For example, Yandong Ma *et. al*[12] have shown that band gaps of GaX monolayers can be tuned by applied mechanical deformation. Xianxin Wu *et. al.*[13] have achieved a long range ferromagnetic coupling between vacancy-induced local moments. Yujie Bai *et. al.*[14] have theoretically illustrated that Tellurium doped GaS monolayers are best suited for the photocatalytic water splitting. H. R. Jappor *et. al.*[15] have shown the indirect to direct band gap tunability of group-III mono-chalcogenides by means of hetero-structuring and predicted Van Hove singularity near valence band edges.

The role of impurity states in semiconducting optoelectronic devices is well known. Mg doping is very common in many semiconductors including GaN to achieve p-type conductivity. Yuting Peng et. al.[16] have shown that Mg impurity can offer effective p-type carriers in the GaSe nanosheets. Possibilities shown by results motivated us to look for the tunability of properties of GaS via group II (C, Si, Ge) and group IVa (Be, Mg, Ca) doping which has not been explored so far. The objectives of the present study have been to study the *i)* structural stability of doped GaS; *ii)* modulations in electronic band structure of GaS monolayer; *iii)* tunability in optical properties including absorption spectra, electron energy loss spectra and static dielectric constant; and *iv)* dopant dependent tunneling current characteristics of GaS monolayer.

## 2. Computational Details

First principle calculations were performed within the framework of DFT using projector augmented-wave (PAW) method[17] as implemented in VASP code[18]. The exchange and correlation energies were treated within the generalized gradient approximation (GGA) with in the Perdew-Burke-Ernzerhof parameterization[19]. The Pristine GaS monolayer was modeled by four atoms with two atoms of each species (i.e. Gallium and Sulphur atoms) in unit cell. A vacuum of ~25 Å along the *z*-direction was introduced to minimize the interaction between the periodic images of 2D system. For the geometry optimization and the optical property calculation, a kinetic-energy cutoff of 400 eV is used. The Brillouin zone was sampled by $\Gamma$ centered *k* mesh of 15x15x1. The self-consistency for electronic steps was carried out with convergence threshold of $10^{-6}$ eV. The structures were relaxed using standard conjugate gradient technique until the force on each atom reaches < 0.01 eV/Å. The electronic density of states (DOS) was calculated using Gaussian smearing for each atomic orbital with broadening parameter of 0.1 eV. The frequency dependent dielectric properties have been obtained after getting electronic ground state with all the unoccupied states in the conduction band.

## 3. Results and Discussion

### 3.1 Structural and Electronic Properties

It has been previously reported that GaS can exist in α as well as in β form[16]. Both of the forms are found to be energetically stable[20]. We considered an α type GaS monolayers in which Sulfur atom is on top of Sulfur atom and Gallium atom on top of Gallium atom as shown in figure 1. The structural parameters; bond length ($R_{Ga-X}$; X=C, Si, Ge, Be, Mg, Ca), bond angle ($\theta_{X-Ga-X}$) and formation energy ($E_f$) are presented in table 1. Our calculated lattice constant of 3.64 Å for GaS monolayer is in good agreement with the available experimental value of 3.57 Å[21] and previously reported theoretical values of ~3.62[16, 22]. Furthermore, the interatomic distance ($R_{Ga-Ga}$) is also in good agreement with the earlier reports[16, 23]. Our calculated results show that $R_{Ga-X}$ distance decreases considerably with C doping and increases with Ge, Mg, Ca with doping. Bond angles (θ) are also found modulated with doping of these elements. This is attributed to the comparatively small (large) atomic size of C (Ge, Mg, Ca) atoms as compared to Ga atom.

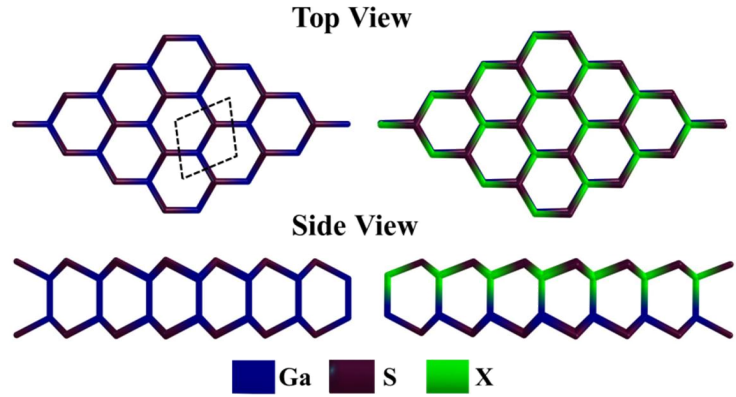

**Figure 1** Top and side view of the 4x4 GaS and $X_{0.5}Ga_{0.5}S$ monolayer (X= C, Si, Ge, Be, Mg, Ca). The dashed line represents the unit cell within the supercell.

To ensure the stability of doped systems we have calculated the formation energy ($E_f$) as

$$E_f = E_{GaS} - (E_{Ga_{0.5}S} + E_X)$$

Where, $E_{GaS}$ is total energy of pristine GaS monolayer, $E_{Ga_{0.5}S}$ is total energy of GaS with 50% of Ga concentration and $E_X$ is total energy of dopant as obtained from converged VASP runs. The negative values of formation energy (table 1) indicate the doping to be exothermic in nature. The formation energy is very large (small) for the case of C (Mg) doped GaS. The large (small) value of formation energy is attributed to large (small) electronegativity difference of C (Mg) with respect to Ga. The Formation energy of Mg and Ca doped GaS differ by 0.60 eV. This small difference is due to a small difference in electronegativity of Mg and Ca dopant.

**Table 1:** Calculated structural parameters (bond length, $R_{Ga-X}$ where, X=C, Si, Ge, Be, Mg, Ca; bond angle, $\theta_{X-Ga-X}$) and formation energy $E_f$ for doped GaS monolayer.

| System | $R_{Ga-X}$ (Å) | | Bond Angle (degree) | | Formation Energy (eV) |
|---|---|---|---|---|---|
| | our | others | $\theta_{X-Ga-Ga}$ | $\theta_{Ga-Ga-X}$ | $E_f$ |
| GaS | 2.47 | 2.33[a], 2.46[b], 2.36[c] | 117.43 | 117.51 | - |
| $C_{0.5}Ga_{0.5}S$ | 1.98 | - | 127.33 | 110.62 | -8.29 |
| $Si_{0.5}Ga_{0.5}S$ | 2.42 | - | 111.65 | 117.31 | -5.94 |
| $Ge_{0.5}Ga_{0.5}S$ | 2.65 | - | 118.57 | 117.77 | -5.16 |
| $Be_{0.5}Ga_{0.5}S$ | 2.48 | - | 97.25 | 117.75 | -6.99 |
| $Mg_{0.5}Ga_{0.5}S$ | 2.81 | - | 114.53 | 117.68 | -4.06 |
| $Ca_{0.5}Ga_{0.5}S$ | 3.22 | - | 126.54 | 118.18 | -4.66 |

[a]Ref[16], [b]Ref[23a], [c]Ref[23b]

In order to get an insight about the modulation in formation energy we have analyzed the charge density difference profiles. The charge density difference has been calculated by subtracting the total charge density of pristine GaS monolayer ($\rho_{GaS}$), from sum of charge density of GaS with 50% Ga concentration ($\rho_{Ga_{0.5}S}$) and charge density of dopant ($\rho_X$) as following

$$\Delta\rho = \rho_{GaS} - (\rho_{Ga_{0.5}S} + \rho_X)$$

The obtained charge density difference plots are presented in figure 2. Red and green color depicts charge accumulation and depletion respectively. The GaS monolayer gets considerably polarized with redistribution of charges. Most of the charge redistribution occurs on the upper layer of GaS due to its proximity to dopant. The polarization is more pronounced for the case of Mg and Ca doping giving signature of strong bonding between dopant and S atom. Charges get depleted from S atom and accumulate between X-S and X-Ga bonding region while for the case of Mg and Ca doping charges get depleted from the X (=Mg, Ca) atom and accumulates on S atom. The shared electrons of Mg and Ca are taken away by S and accumulating on S. This is because of greater electronegativity of S as compared to Mg and Ca resulting in lower formation energy.

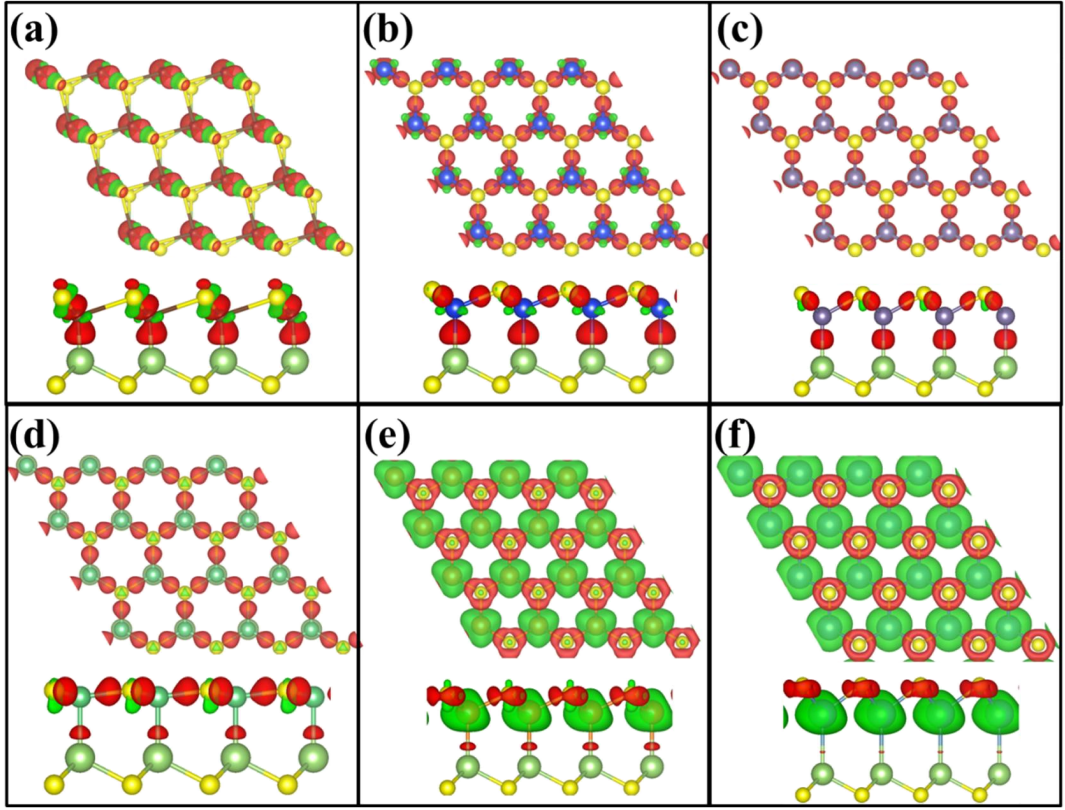

**Figure 2** Side and top views of charge density difference profiles for **a)** $C_{0.5}Ga_{0.5}S$, **b)** $Si_{0.5}Ga_{0.5}S$, **c)** $Ge_{0.5}Ga_{0.5}S$, **d)** $Be_{0.5}Ga_{0.5}S$, **e)** $Mg_{0.5}Ga_{0.5}S$, **f)** $Ca_{0.5}Ga_{0.5}S$. Isosurface value is set at $1 \times 10^{-2}$ e/Å$^{-3}$.

Furthermore, the electronic band structures are calculated along the high-symmetry Γ-M-K-Γ direction of the Brillouin zone. The atom projected electronic band structure and corresponding orbital projected Density of States (DOS) is presented in figure 3. Atom projected electronic band structure shows pristine GaS to be semiconductor with an indirect band gap as marked with $E_g^A$ and $E_g^B$ in figure 3. Note that difference between the band gaps $E_g^A$ and $E_g^B$ is very small of the order of 0.1 eV. Our calculated value of band gap of 2.40 eV is underestimated by an amount of 0.20 eV as compared to experimentally observed value of 2.60 eV[16]. The underestimation in calculated band gap as compared to experimentally measured values is attributed to well-known deficiency of DFT level of theory. Atom projected band structure reveals that valance band maximum (VBM) is delocalized over Ga and S while a conduction band minimum (CBM) is localized on the S atom. This is confirmed by orbital projected density of states. The contribution to VBM originates from strongly hybridizing Ga-$p_z$ and S-$p_z$ orbitals. The contribution to CBM is significantly from S-$p_z$ orbitals.

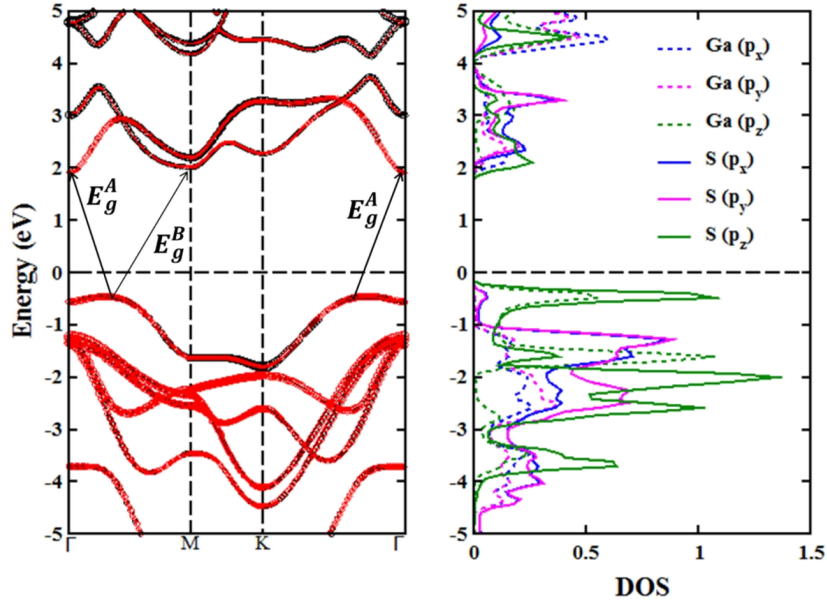

**Figure 3**: Atom projected electronic band structure and corresponding orbital projected density of states for pristine GaS monolayer. Fermi energy has been set at 0 eV.

Furthermore, the doping of GaS results semiconductor to metallic transition due to significant shift in energy levels as can be seen in figure 4. To gain an insight into significant shift in energy levels we looked at the modulations in Fermi energy. It has been observed that for C, Si and Ge doping pushes Fermi level into the conduction band while for Be, Mg and Ca doping uplift the Fermi level into the valance band with respect to Fermi level of pristine GaS. This shifting results in p-type and n-type doping effects as can be seen clearly in projected density of states (figure 5). The energy levels in the vicinity of Fermi energy are dominated by *X-p* and *S-p* orbitals. A similar characteristic features can been seen in figure 4. The finite density of states at Fermi energy confirms the metallic character of doped GaS. Since shifting in energy bands give clear signatures of energy level offset. Therefore, we have analyzed the valance and conduction band offset. The relative alignment of valance band and conduction band can be quantified in terms of the energy level offset (band offset)[24]. Experimentally, the STM and µ-XPS techniques are capable of measuring band offsets[25]. We define the valance band offset (VBO); $\Delta E_V = E_V^{GaS} - E_V^{X_{0.5}Ga_{0.5}S}$ and conduction band offset (CBO); $\Delta E_C = E_C^{GaS} - E_C^{X_{0.5}Ga_{0.5}S}$. Calculated

VBO and CBO have been tabulated in table 2. The negative (positive) values indicate the upward (downward) shift in valance band maxima (conduction band minima). Doping of C, Si and Ge indicates p-type doping effect while Be, Mg and Ca exhibit n-type doping effect. This offset in energy bands is principle reason for semiconductor to metallic transition upon doping in all the cases.

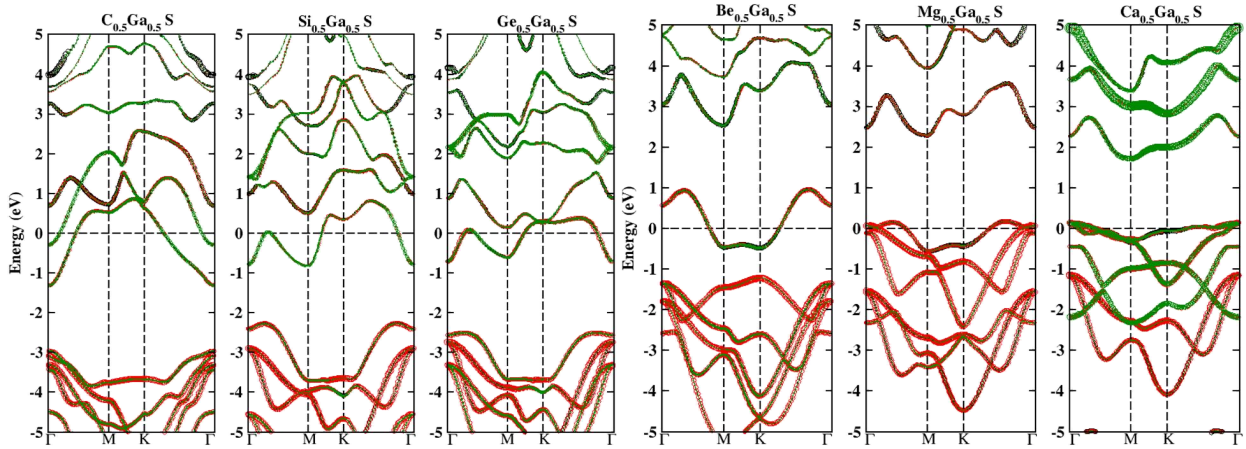

**Figure 4**: Atom projected electronic band structures for doped GaS monolayer. Fermi energy has been set at 0 eV. Black, red and green color represent contributions from Ga, S and X (=C, Si, Ge, Be, Mg, Ca) atoms respectively.

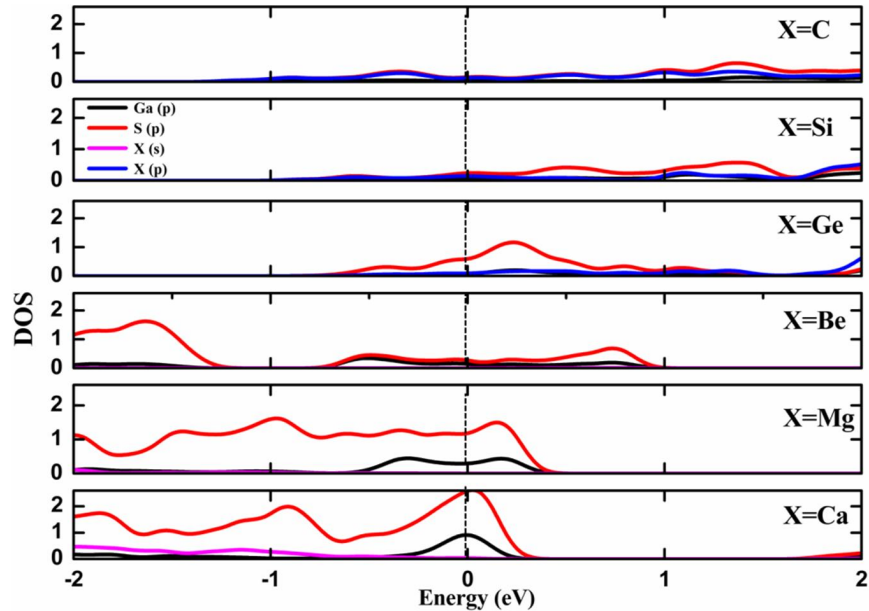

**Figure 5** Projected density of states for doped $X_{0.5}Ga_{0.5}S$ monolayer.

Table 2 Valance band and conduction band offset (in eV) for doped GaS monolayer with reference to valance band and conduction band energy levels of pristine GaS.

| System | Valance Band Offset | Conduction Band Offset |
|---|---|---|
| $C_{0.5}Ga_{0.5}S$ | 2.48 | 3.22 |
| $Si_{0.5}Ga_{0.5}S$ | 1.78 | 2.73 |
| $Ge_{0.5}Ga_{0.5}S$ | 2.04 | 2.66 |
| $Be_{0.5}Ga_{0.5}S$ | -1.40 | -0.62 |
| $Mg_{0.5}Ga_{0.5}S$ | -0.64 | -0.39 |
| $Ca_{0.5}Ga_{0.5}S$ | -0.60 | 0.16 |

## 3.2 Optical Properties

Furthermore, effect of doping on dielectric properties of GaS has been investigated. Dielectric properties have been calculated in the energy range from 0 to 15 eV. The Dielectric response has been recorded for electric vector E perpendicular to c axis (E⊥c) and parallel to c axis (E∥c). The complex dielectric function can be expressed as sum of real ($\mathcal{E}_1$) and imaginary part ($\mathcal{E}_2$). First imaginary part of dielectric function ($\mathcal{E}_2$) is evaluated using following relation[26]

$$\varepsilon_2^{\alpha\beta} = \frac{4\pi^2 e^2}{\Omega} \lim_{q \to 0} \frac{1}{q^2} \sum_{c,v,k} 2\omega_k \delta(\varepsilon_{ck} - \varepsilon_{vk} - \omega) \langle u_{ck+e_\alpha q} | u_{vk} \rangle \langle u_{ck+e_\beta q} | u_{vk} \rangle$$

where, indices $\alpha$ and $\beta$ represent cartesian components, $v$ and $c$ is used to represent valance and conduction band and $k$ is used for k-points of the brillioun zone; $e_\alpha$ and $e_\beta$ are unit vectors along x, y and z direction; $\varepsilon_{ck}$ and $\varepsilon_{vk}$ represent energy of valance and conduction band and $u_{vk}$ ; and $u_{uk}$ refers to cell periodic part of orbitals. Once $\mathcal{E}_2$ is calculated, $\mathcal{E}_1$ is obtained using Kramer's-Kronig relations. The complex dielectric function for in-plane polarization (E⊥c) and out-of-plane polarization is obtained using following relations

$$\varepsilon^\perp(\omega) = \frac{\varepsilon^{xx}(\omega) + \varepsilon^{yy}(\omega)}{2}$$

and
$$\varepsilon^\parallel(\omega) = \varepsilon^{zz}(\omega)$$

where $\varepsilon^{xx}(\omega)$, $\varepsilon^{yy}(\omega)$ and $\varepsilon^{zz}(\omega)$ are diagonal elements of the dielectric matrix. Electron energy loss (EEL) function has been calculated from dielectric functions as

$$EEL = Im\left\{\frac{1}{\varepsilon(\omega)}\right\} = \frac{\varepsilon_2(\omega)}{\varepsilon_1^2(\omega) + \varepsilon_2^2(\omega)}$$

Figure 6 and 7 depicts the calculated $\varepsilon_1$, $\varepsilon_2$ and *EEL* spectra for pristine and doped GaS. Peak positions in imaginary part ($\varepsilon_2$) and loss spectra of pristine and doped GaS have been tabulated in table 3. For pristine GaS two pronounced peaks at 3.72 eV and 6.56 eV for out-of-plane polarization and a single peak at 4.95 eV for in-plane polarization. A single peak at 4.05 eV was observed by H.R. Japoor *et. al.*[23a] in the neighborhood of our observation. Also, absorption began in visible region and gets pronounced in ultraviolet region. The imaginary part ($\varepsilon_2$) of dielectric function exhibit distinct spectral feature for all dopants in both the polarizations. The $\varepsilon_2$ show a significant blue shift in UV-region for both the polarizations. New peaks appeared in visible region indicating doped GaS as UV/Vis active material.

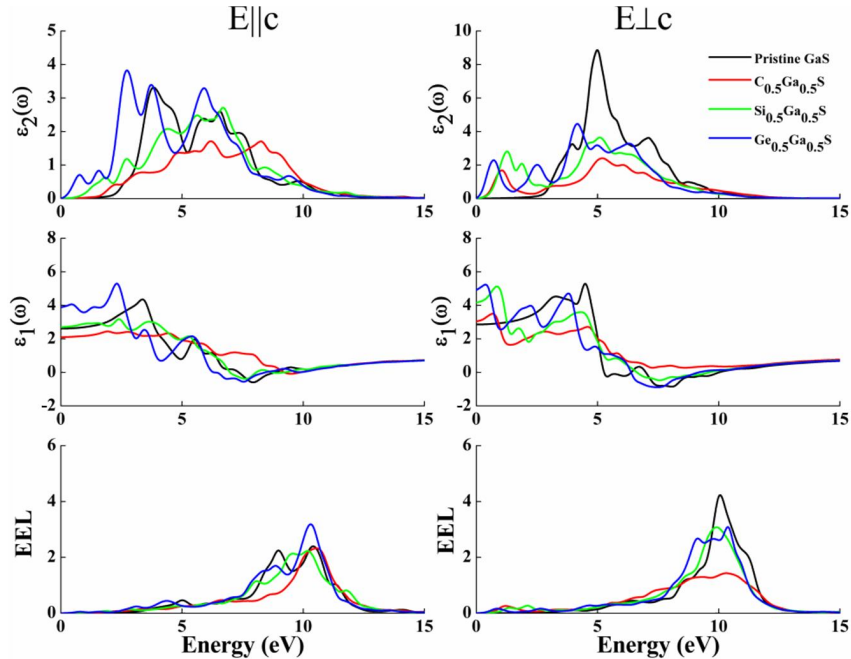

**Figure 6** Imaginary parts ($\varepsilon_2$) and real part ($\varepsilon_1$) of dielectric function and electron energy loss (EEL) spectra for C, Si and Ge doped GaS for E‖c and E⊥c axis.

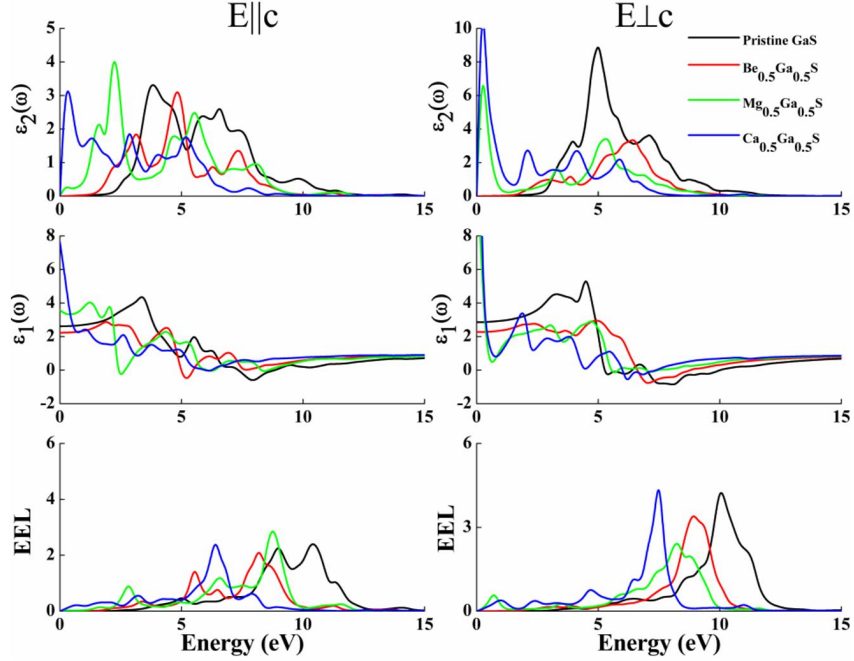

**Figure 7** Imaginary part ($\varepsilon_2$) and real part ($\varepsilon_1$) of dielectric function and electron energy loss (EEL) spectra for Be, Mg and Ca doped GaS for E||c and E⊥c axis.

Furthermore, in real part of dielectric function ($\varepsilon_1$) negative values (figure 6 and 7) have been observed between 5 eV to 11 eV (UV region). The negative values suggest the metallic nature of doped GaS in this region. The point where negative value of $\varepsilon_1$ starts to be positive corresponds to collective excitations of electrons. The value of real part of dielectric function at zero energy (i.e. $\varepsilon_1$ (0)) gives the measure of static dielectric constant ($\varepsilon_s$). The value of $\varepsilon_s$ for pristine GaS is found to be 2.62 and 2.85 for E||c and E⊥c respectively. Our calculated vale for E||c is consistent with the earlier reported value of 2.68[23a]. The static dielectric constant generally increases for the doped GaS as compared to pristine for both E||c and E⊥c. This increased value of $\varepsilon_s$ is attributed to semiconductor-to-metallic transition which occurs upon doping.

The role of plasmons is very important in metals and semiconductors. The calculated EELS show insignificant modulation in low energy region while in high energy region plasmonic features are significantly modulated. A plasmonic feature in low energy region (at 2.78 eV) appears for Be doped GaS. This corresponds to excitations of π-electrons. A strong plasmonic feature is observed in the high energy region (~10 eV) for both the polarizations (figure 6 and 7). The excitations in the high energy region are due to collective excitations of π and σ electrons.

EEL spectra of GaS shift towards high energy region (red shift) with Be, Mg and Ca doping in both the polarizations. No significant shift in EEL spectra is observed for C, Si and Ge doped GaS. All the prominent peaks in EEL spectra have been presented in table 3. The values of EEL correspond to the energy value where real part of dielectric function cross the x-axis from negative values. These shifts in EEL spectra and appearance of new low energy plasmons may be useful for optical bio-sensing.

**Table 3** Peak positions in imaginary part of dielectric function ($\varepsilon_2$), loss spectra, and static dielectric constant ($\varepsilon_s$) for pristine and doped GaS monolayers.

| System | $\varepsilon_2$ (eV) | | Loss Spectra (eV) | | $\varepsilon_s$ | |
|---|---|---|---|---|---|---|
| | E∥c | E⊥c | E∥c | E⊥c | E∥c | E⊥c |
| GaS | 3.72, 6.56 | 4.95 | 8.94, 10.42 | 6.22, 10.04 | 2.62 | 2.85 |
| | | | | | | 2.68[a] |
| $C_{0.5}Ga_{0.5}S$ | 6.15, 8.22 | 1.04, 5.12 | 10.54 | 10.25 | 2.10 | 3.05 |
| $Si_{0.5}Ga_{0.5}S$ | 1.81, 2.61, 4.35, 5.63, 6.65 | 1.25, 1.85, 5.07 | 8.04, 9.39, 10.17 | 9.96 | 2.71 | 4.18 |
| $Ge_{0.5}Ga_{0.5}S$ | 0.75, 1.47, 2.70, 3.68, 5.93, 9.32 | 0.74, 2.57, 4.18, | 8.85, 10.26 | 9.11, 10.42 | 3.89 | 4.90 |
| $Be_{0.5}Ga_{0.5}S$ | 3.12, 4.74, 7.33 | 6.43 | 5.50, 8.13 | 8.85 | 2.24 | 2.29 |
| $Mg_{0.5}Ga_{0.5}S$ | 1.64, 2.19, 4.61, 5.50, 8.05 | 0.28, 3.33, 5.33 | 2.78, 6.56, 8.73 | 5.89, 8.32 | 3.66 | 13.00 |
| $Ca_{0.5}Ga_{0.5}S$ | 0.28, 1.30, 2.87, 5.16, 7.79 | 0.24, 2.02, 4.14, 5.88 | 6.39 | 4.63, 7.42 | 7.61 | 27.46 |

[a]Ref[23a]

### 3.3 Transport Properties

Since, it is well known that the work function of any material plays an important role in device performance. Work function has been calculated for pristine and doped GaS monolayer. The work function is calculated as[27]

$$\Phi = E_{vac} - E_F$$

where $E_{vac}$ and $E_F$ are vacuum and Fermi levels respectively. The calculated values of work function have been tabulated in table 4. The work function of pristine GaS is 3.50 eV. Our calculated work function of GaS is smaller than that of graphene (4.51 eV)[28] and MoS$_2$ (5.11 eV)[29]. Mg doping resulted in increase of work function up to 6.68 eV as compared to pristine GaS. Note that doped GaS has work functions located in the vicinity of pristine graphene and MoS$_2$. The increase in work function of GaS with doping is attributed to large electronic redistributions (figure 2) which resulted in modulation in electrostatic potential.

Furthermore, modulations in electronic energy levels in the vicinity of Fermi level with doping are expected to modify electron transport properties as compared to pristine monolayer. The expected modulation is quantified in terms of quantum ballistic conductance. The quantum ballistic conductance can be determined by the number of electronic energy levels crossing the Fermi level with each having conductance of G$_0$. Therefore, n-bands crossing the Fermi level will have conductance equal to $nG_0$. We find the conductance of 4G$_0$ for doped GaS except for Be-doping.

**Table 4:** Calculated work function ($\Phi$) and quantum ballistic conductance for pristine and doped GaS monolayer.

| System | $\Phi$ (eV) | Conductance |
| --- | --- | --- |
| GaS | 3.50 | - |
| C$_{0.5}$Ga$_{0.5}$S | 4.61 | 4G$_0$ |
| Si$_{0.5}$Ga$_{0.5}$S | 3.91 | 4G$_0$ |
| Ge$_{0.5}$Ga$_{0.5}$S | 4.17 | 4G$_0$ |
| Be$_{0.5}$Ga$_{0.5}$S | 5.40 | 2G$_0$ |
| Mg$_{0.5}$Ga$_{0.5}$S | 6.04 | 4G$_0$ |
| Ca$_{0.5}$Ga$_{0.5}$S | 6.68 | 4G$_0$ |

Furthermore, the expected modulations in transport properties are captured by calculating current voltage characteristics with in the STM like setup. In this model setup, $Au_{13}$ in icosahedral geometry has been used to simulated STM-like tip. Transport characteristics are obtained within Bardin, Tersoff and and Hamman (BTH) formalism[30]. The tip and sample are kept at sufficient gap of 4Å to mimic nonbonding configuration. It is noticeable that the magnitude of current depends on the tip sample separation although the characteristic feature remains the same. The calculated tunneling characteristics for the pristine and doped GaS have been presented in figure 8. The pristine GaS monolayer is semiconductor while the doped GaS shows ohmic behavior for low forward bias of ≤ 0.2 V. For reverse bias asymmetric characteristics for doped GaS monolayer appear with modulation of current. This can be understood from density of states as the conduction channels are governed by the available states around the Fermi level. For example, $Be_{0.5}Ga_{0.5}S$ has very small tunneling current for a given bias. This is attributed to low availability of channels for conduction as can be seen from density of states around Fermi energy (figure 5). On the other hand $Ge_{0.5}Ga_{0.5}S$ has comparatively large magnitude of current at a given bias due to larger number of available conduction channels. These interesting features underline the potential of doped GaS for device applications. Note that the electron transport properties using *Au* was investigated theoretically here for providing guidance for metallic contact for GaS based devices.

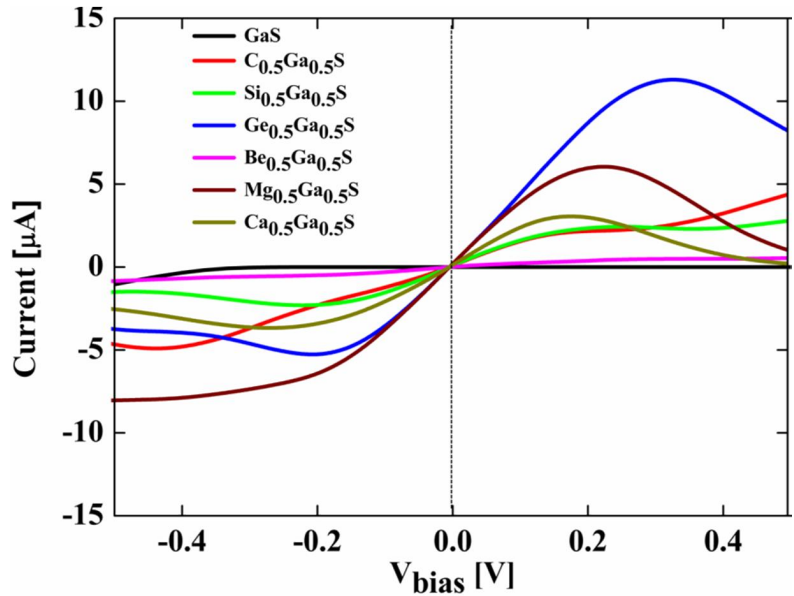

**Figure 8.** Tunneling current characteristics for pristine and doped GaS monolayer.

## 4. Summary

Stability, electronic, optical and transport properties of doped GaS monolayer are investigated using density functional theory. The group II and group IVa doped GaS is found to be energetically stable. The pristine GaS monolayer is an indirect band gap semiconductor and doping of group II and group IVa elements makes GaS metallic. Imaginary part of dielectric function in doped GaS show new peaks in visible region indicating its usefulness in photo-diodes, photo-catalysts, and photo-detectors. Calculated work function of pristine and doped GaS located in the vicinity of graphene and $MoS_2$. The modulation in electronic energy levels is confirmed by tunneling current characteristics showing ohmic-like characteristics for doped GaS monolayer. We believe that the results of the present study will hold a key to extend the functionality of GaS monolayer as a potential candidate for applications in optoelectronic devices at nanoscale.

**Acknowledgements**

Helpful discussions with Dr. Kulvinder Singh, Dr. Priyanka Sharma and Dr. Anshu Sharma are highly acknowledged. CVRAMAN; high performance computing cluster at Himachal Pradesh University was used in obtaining the results presented in this paper.